# Two Different Experiments with the Rope-Attached Sphere by Using Arduino

Atakan Çoban[1], Ercüment Akat[1] and Ahmet Cem Erdoğan[1]
[1]Faculty of Arts and Sciences, Department of Physics, Yeditepe University, Ataşehir, Istanbul, Turkey

## Abstract

Arduino microcontrollers are electronic devices that can be used in physics experiments, and they are both affordable and readily available in comparison to other experimental sets. Students in high school and college can acquire valuable skills through the usage of Arduino in STEM education applications. In this study, two different physics experiments were conducted with a single, easy to develop, experimental setup by using Arduino, a force sensor and a rope-attached metallic sphere. First experiment, involved several analyses such as period, rope tension, and energy conservation during simple harmonic motion. In the second experiment, the metal ball was released from the rope's connecting point, and the impulse acting on it between the times it was released and when it returned to equilibrium was measured. Analyzes performed are in very precise force and time intervals and the results of all analyzes are consistent with the theoretical values. Educational activities in this direction can both create maximum gains for students in STEM fields in limited education periods and contribute to equality of opportunity in education due to its economic nature.

## 1. Introduction

Recent technological advances have opened up new ways to teach and learn the skills that will be most important in the 21st century[1]. STEM Education, which is based on integrating Science, Technology, Engineering, and Math, has recently become a part of how technology is changing in the modern world[2]. STEM education's technology part is often done with Arduino microprocessors, which can measure and control many parameters and ideas by connections with compatible sensors[3,4]. These sensors can measure distance, speed, acceleration, force, sound, light, temperature, current, potential difference and many more parameters. Using Arduino microprocessors, researchers have recently done a lot of research on how to teach different physical laws and ideas[5-7]. In this study a force sensor and an Arduino device are used. A metallic sphere is connected to the tip of the force sensor by a rope. Two different activities are realized in three different physical topics by the use of this experimental system. In the first activity the sphere and the tight rope are held at a certain initial angle from vertical and the sphere is released to oscillate as a simple pendulum to analyse the subsequent motion. In the second activity the sphere is set free from a certain height and some analyses are carried out on it about the total impulse acting on it until it stops. With the help of a single system that can be easily and cheaply prepared the above mentioned studies include basic applicational improvements in two different elementary topics within the context of physics.

## 2. Theory

In this study the cases such as the buoyancy force, simple pendulum and the falling motion of the sphere attached to the end of the rope are examined. In this part of the study, basic principles and mathematical equations will be given by using the book information[8,9].

*Simple pendulum*

If a massive sphere (called the "*bob*") attached to the end of a massless rope is displaced even slightly from the vertical equilibrium position it exhibits oscillations in a vertical plane, by sweeping almost equal angles on both sides, in cases where the air friction can be neglected. This motion is called "*simple harmonic motion (SHM)*" and this system is called "simple *pendulum*". The time elapsed for a full round trip is called the "*period*" of the motion which is calculated by the expression,

$$T' = 2\pi\sqrt{\frac{l}{g}} \qquad (1)$$

$T'$ is the period, $l$ is the length of the rope and $g$ is the local gravitational acceleration in equation 1. The full period of the SHM of the sphere can be considered to consist of four parts. i- At time t=0 it is displaced aside from the vertical by a small angle $\theta$ (<15) and is released. ii- At $t=T'/4$ it passes by the vertical equilibrium position at its maximum speed. iii- At $t=T'/2$ it reaches the symmetrically opposite position where it momentarily comes to rest. iv- At $t=3T'/4$ again it passes by the vertical equilibrium position at its maximum speed but now in the opposite sense. v- Finally at $t=T'$ it completes a full period and it momentarily comes to rest again. The first quarter of the motion can be seen from Fig. 1.

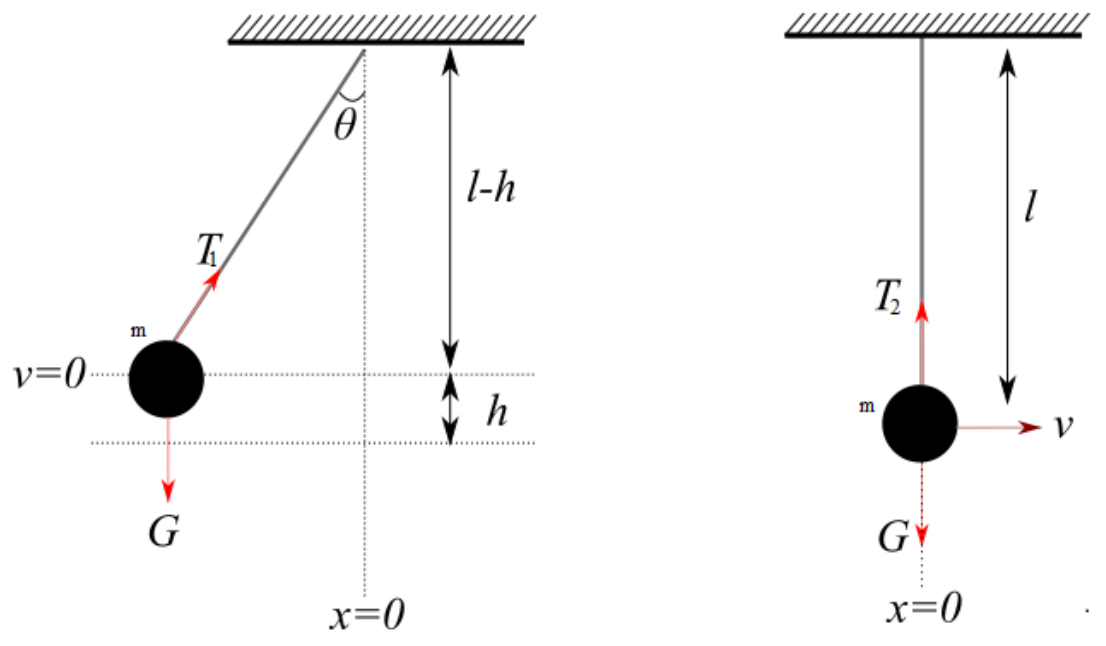

**Fig. 1.** The diagram for the first quarter of the simple harmonic motion of period $T'$, executed by a simple pendulum of length *l* and mass *m*.

During the harmonic motion the tension force *T* in the rope varies according to the angular position of the bob. The tension *T* reaches its maximum value $T=T_{max}$ at the vertical position where the linear velocity of the bob reaches its maximum possible value. This maximum value can be found by assuming that the bob executes a short-range circular motion. At this position the difference between the two opposite forces, namely, the vertically up tension *T* in the string and the vertically down weight of the bob supplies the necessary centripetal acceleration for the bob to exhibit a circular motion, at least locally. The equation of motion for the bob at this moment is,

$$T_{max} - mg = m\frac{v^2}{l} \qquad (2)$$

where $l$ is the length of the rope, $m$ and $v$ are the mass and velocity of the bob respectively. The positions at which the tension $T$ takes its minimum possible value ($T_{min}$) are the two limits where it comes to rest momentarily. Since the zero value of the potential energy can be chosen anywhere it is appropriate to set y=0 at the lowest position of the motion. Therefore the height of the bob at rest y=h corresponds to the position where the mechanical (total) energy is purely potential and the bob has no kinetic energy. From the conservation of energy this height $h$ can be determined from

$$0 + mgh = \frac{1}{2}mv^2 + 0 \qquad (3)$$

From the figure it is seen that at this position the vertical distance between the level of the bob and that of the upper (fixed) end of the rope is $l$-h and the angle can be found from the trigonometric expression

$$\cos\theta = \frac{l-h}{l} \qquad (4)$$

As can be seen from the free-body diagramme the tension at the highest position is $T_{min}$=mgcos$\theta$, and it has both a horizontal and a vertical component. The force sensor used in this work can take measurements only in one dimension and the vertical component of the tension is measured during the simple harmonic motion. The compatibility between the experimental and theoretical values of the vertical component of the tension is calculated by the equation

$$T_y = mg\cos^2\theta \quad (5)$$

*The fall of a sphere attached to a rope*

The objects released from a certain height execute freely-falling motion with the gravitatonal acceleration $g$. If an object is attached to the free end of a massless and inextensible rope of length $l$ and is then released it falls under the effect of its weight $mg$.

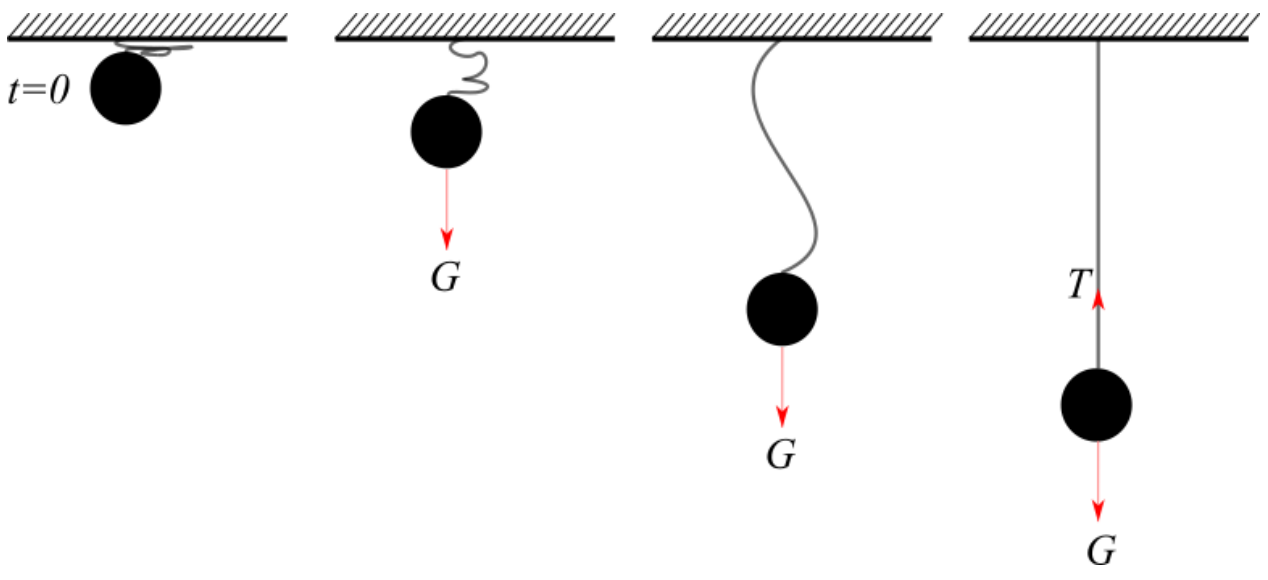

**Fig. 2.** The motion of a falling sphere attached to the end of a rope.

It subsequently is pulled up by the tension $T$ in the rope and after a while comes to rest at the end of the rope as shown in Fig. 2. During this motion the net force acting on the body is

$$\vec{F} = m\vec{g} - \vec{T} \quad (6)$$

If a net force $F$ acts on an object during a time interval Δt, then it creates an impulse $I$ that can be found from

$$\vec{I} = \vec{F}(\Delta t) \qquad (7)$$

As can be seen from the equation the impulse is a vector quantity and is always collinear with the force. On the other hand, the impulse $I$ is equal to the change in the momentum $\vec{P} = m\vec{v}$, given by the expression

$$\vec{I} = \vec{F}(\Delta t) = m(\Delta \vec{v}) \qquad (8)$$

For the object which is attached to the free end of the falling rope the change of velocity and the momentum in turn are both zero. Thus, the total impulse created by the net force during the whole motion is zero. In cases where the net force varies as a function of time the area under the "net force vs. time" curve gives the impulse $I$.

In this study the net impulse $I$ during the motion is calculated from the area under the curve "force vs. time" by integration. This integral is calculated by the trapezoidal method for two data points chosen with the expression

$$\frac{1}{2}(F_1 + F_2)(t_2 - t_1) \qquad (9)$$

According to the trapezoidal method[4] in order to calculate the whole integral each and every pair of data should be used and the results should be added. Hence, the integral is calculated by the expression

$$I = \int F dt = \frac{1}{2}\sum_{k=1}^{15}[F_{k-1} + F_k](t_k - t_{k-1}) \qquad (10)$$

## 3. Experimental Setup

The items used in all three analyses include an Arduino UNO, 1 kg load-cell sensor, a metallic sphere of mass 114 g, a piece of rope having length 50 cm and a wooden platform. Also an amplifier of the type HX-711 is used to magnify and transfer the data taken from the sensor. The load-cell sensor is used to weigh the masses in sensible electronic balances. This sensor is used in the study as a force sensor to arrange the codes starting from the relationship between the mass and weight. Furthermore, the vertical component of the tension in the rope holding the metallic sphere is measured under different circumstances. The electronic connections between the sensor, amplifier and Arduino are shown in Fig. 3.

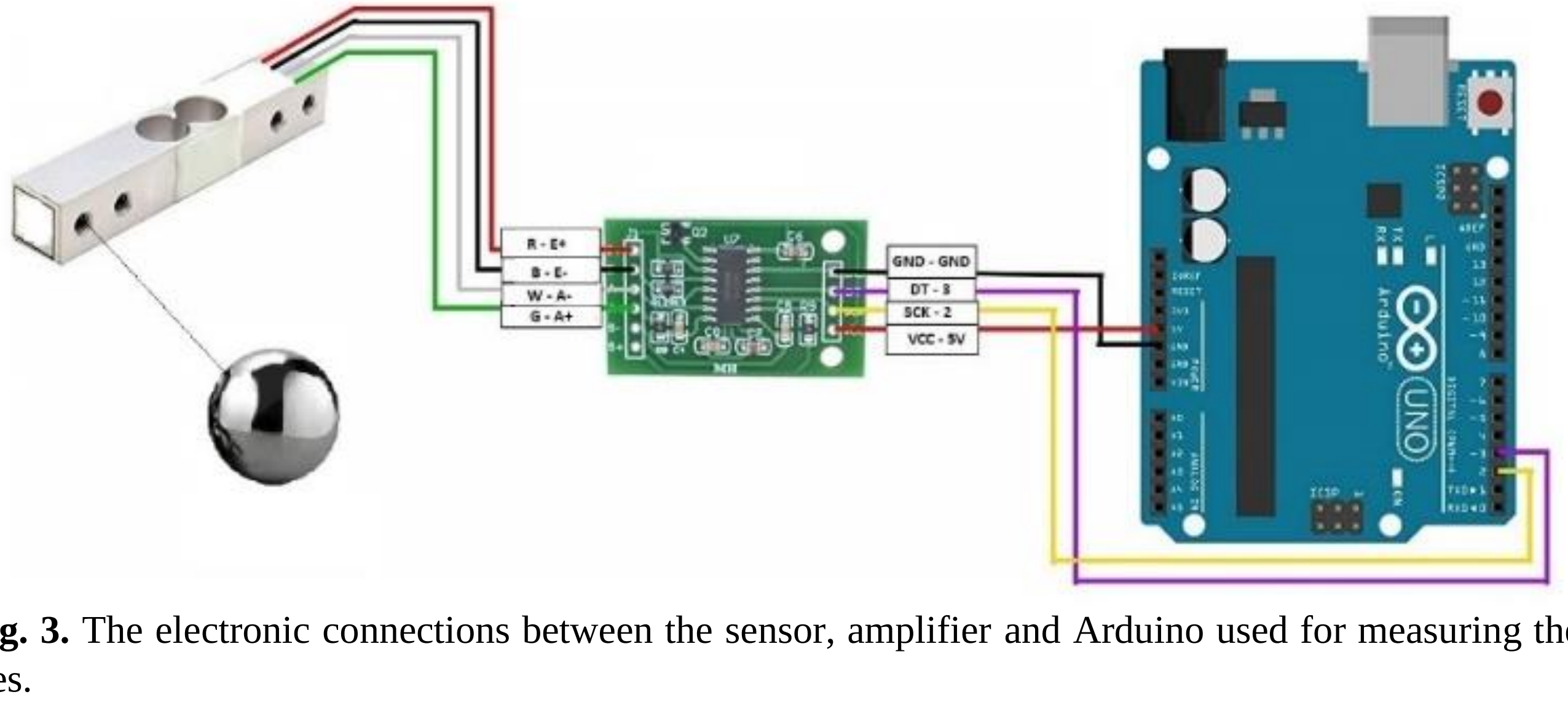

**Fig. 3.** The electronic connections between the sensor, amplifier and Arduino used for measuring the forces.

After the electronic connections are carried out the wooden platform is prepared and the three-dimensional material takes the final form as in Fig. 4.

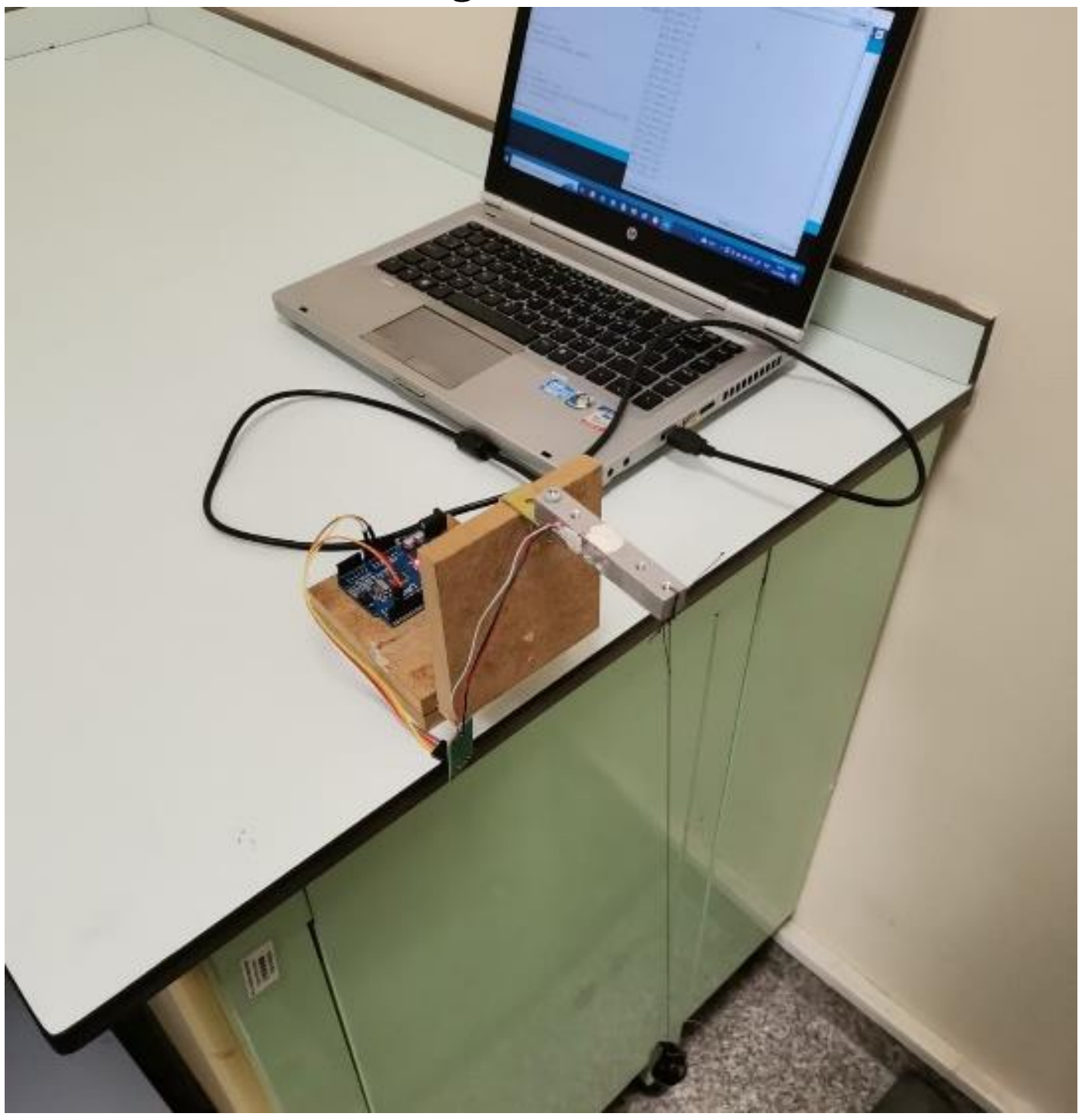

**Fig. 4.** Experimental setup.

The computer programme is written following the preparation of the materials. The programming is carried out using two different sets of codes at the root of which lies the force measurement. In the studies for the determination of the buoyancy analyses and density of the liquid it is aimed to measure the bouyant force acting on the sphere submerged in the liquid and to determine the density by substituting this value in the equation. As for the simple pendulum and falling mass cases the same code is used and the tension in the rope is found as a function of time. Both codes are given in the Appendices.

## 4. Results and Data analysis

*a- Simple pendulum*

The aim of the experiment is to find the period of the simple harmonic motion executed by the simple pendulum and to analyze the tension occurring in the rope. In the context of these analyses the equations for the period and the energy conservation are used. The vertical component of the tension force $T$ is measured during the back and forth oscillations. The positions at which the angle $\theta$ of the rope is maximum are shown in Fig. 5.

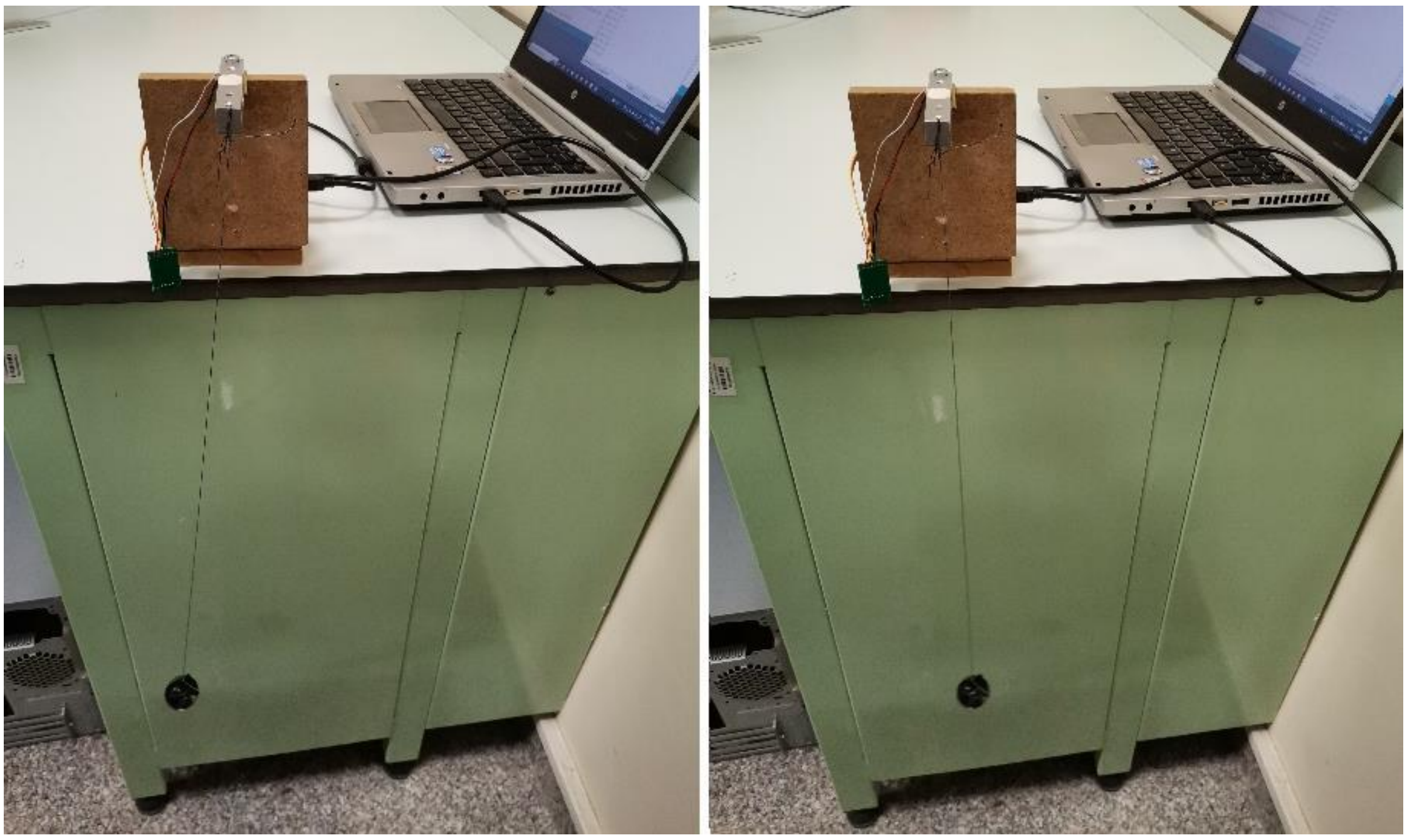

**Fig. 5.** The activities related with the analyses about the simple pendulum.

During the experimental part the first thing is to read the force values when the ball is in equilibrium in vertical situation. The purpose here is to observe that the tension in the rope equals the weight of the sphere when it is at rest: T=mg. Then the bob is pulled aside by an angle not more than $15^0$ from the vertical, because of being limited by the small angle (Taylor) approximation and the bob is then set free. The vertical component of the tension force *T* in the rope is measured. The "tension force vs. time" data are plotted in Fig. 6.

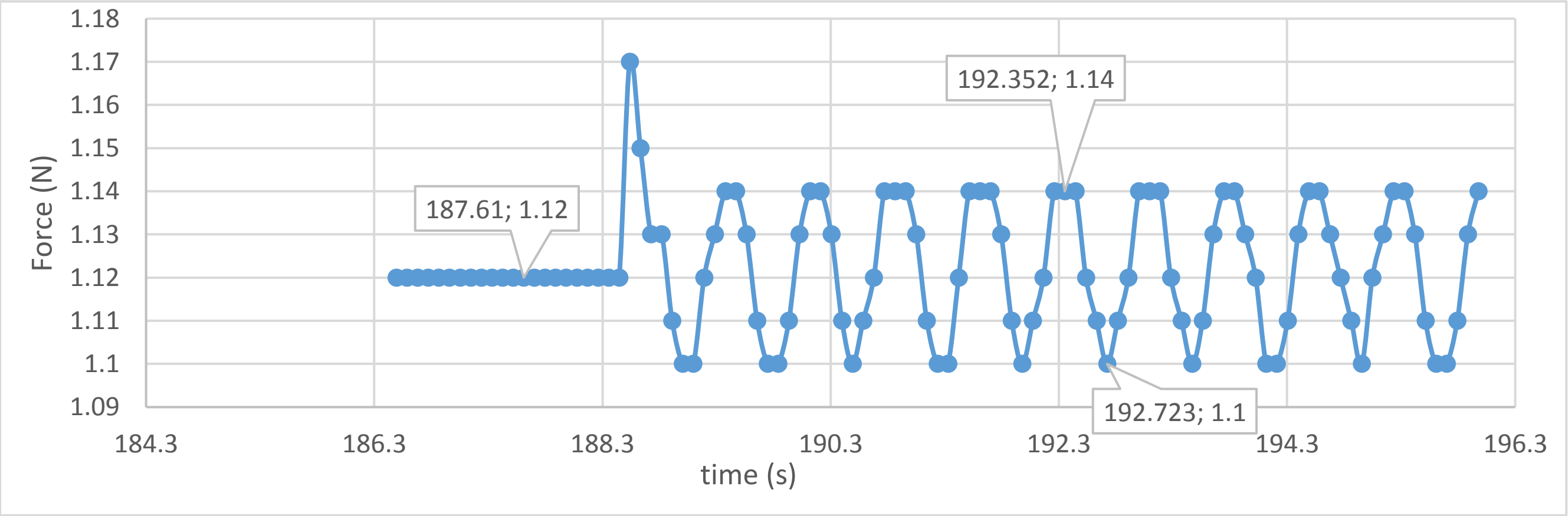

**Fig. 6.** The graph of "tension force vs. time" using the data of the simple pendulum set.

The flat part near the left end of the plot corresponds to the stationary part of the motion where the bob is still at rest vertically. The reading for the tension equals the weight of the bob. Immediately after this region the rope is given a certain small, initial value compatible with Taylor approximation. The transient period in between the flat and sinusoidally oscillating parts belongs to the moment where the rope is given an initial velocity manually and the reason for the fluctuations is external.

This part is deliberately included in the plot in order not to lose fidelity to the data. After the bob is set free a sinusoidal graph is obtained between a certain minimum and a maximum of the oscillations. It is obvious that the vertical component of the tension varies in magnitude as a function of the position of the bob. The maxima in the graph show the values when the bob is passing by the equilibrium length and the minima show those obtained when the bob is at the limits having zero velocity. If the length of the bob is taken as $l$=0.5 m and the gravitational acceleration as g=9.81 m/s$^2$, then the period for the simple pendulum is found to be $T'$=1.42 N by Eq. 1. The time it takes for a pendulum that exhibits SHM to go from the equilibrium (a maximum) to the nearest maximum amplitude (a minimum) is one quarter of the whole period. Four times the distance between a maximum randomly chosen and the minimum immediately after that on the graph gives the period and this value is found to be 1.48 s, with a discrepancy of 4.22% from the theoretical value.

The maximum velocity of the ball while passing by the equilibrium position can be found by Eq. 2. The mass and weight of the bob are m=0.11 kg and G=1.12 N respectively, the maximum value of the tension is $T_{max}$=1.14 N, the length of the rope is $l$=0.5 m and if these values are substituted the velocity of the bob is found to be $v$=0.30 m/s. By using Eq. 3 giving the conservation of energy and using this specified value for the velocity the height of the bob at the limiting position is calculated to be h=4.47 mm. From this result two other values are obtained as $l$-h=0.49 m and cos$\theta$=0.99.

The vertical component of the tension in the rope that is measured by way of the force sensor during the motion as mentioned above is calculated to be $T_y$=1.10 N by Eq. 5 when the angle reaches its maximum amplitude. This value is measured experimentally as 1.10 N as shown in Fig. 6 whereas the previous value of it is found from a set of equations starting from other empirical data. It is seen that the calculated and measured values are the same.

*b- Fall of a sphere attached to a rope*

In the third experimental application a sphere attached to a rope is released from the point where the rope is connected to the sensor. The tension in the rope is recorded as a function of time during the motion from the moment of releasing until it reaches equilibrium. The analyses for the variation in the impulse-momentum on the sphere are carried out by these data. Data gathering in the experimental part is shown Fig. 7.

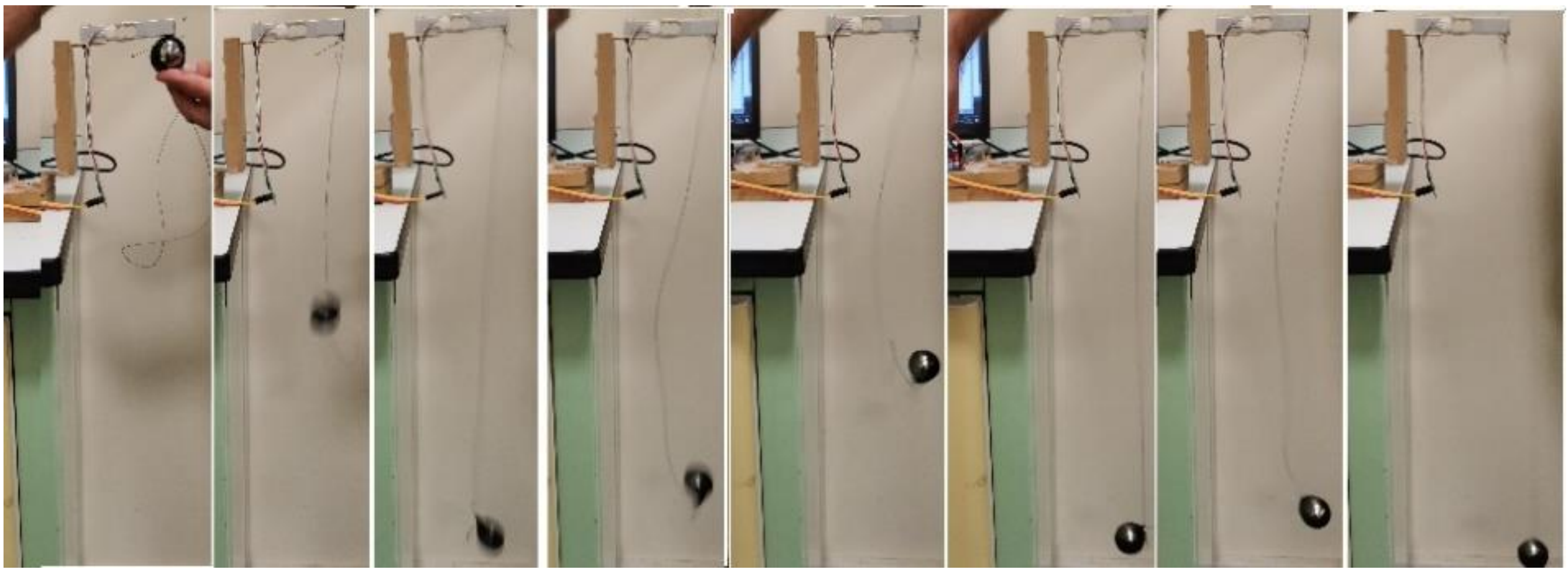

**Fig. 7.** The sphere attached to the end of a rope comes to equilibrium after being set free.

The analyses are based on the data collected from the moment the sphere is released so it is necessary to specify the beginning. Since there is no tightening in the rope while the sphere is falling down the force sensor is expected to read zero for that part.

It is evident that the tension will be nonzero when the sphere reaches its lowest position. It is also obvious that if the sphere is held by hand at a certain height the rope being loose, the tension reading will be zero. In order to specify the moment at which the falling starts it is important not to confuse the data collected before and after this threshold. Therefore the sphere is first held by hand and then released so as to exert a force on the sensor. Hence, some nonzero values are read from the sensor at moments other than the falling part where it is still held. Thanks to this simple operation that facilitates the analysis of data it is quite easy to see the moments at which the falling starts by reading zero from the sensor. While the sphere is held at a certain height and during the period between being set free and coming to stop, the values of the tension are shown in Fig. 8 as a function of time.

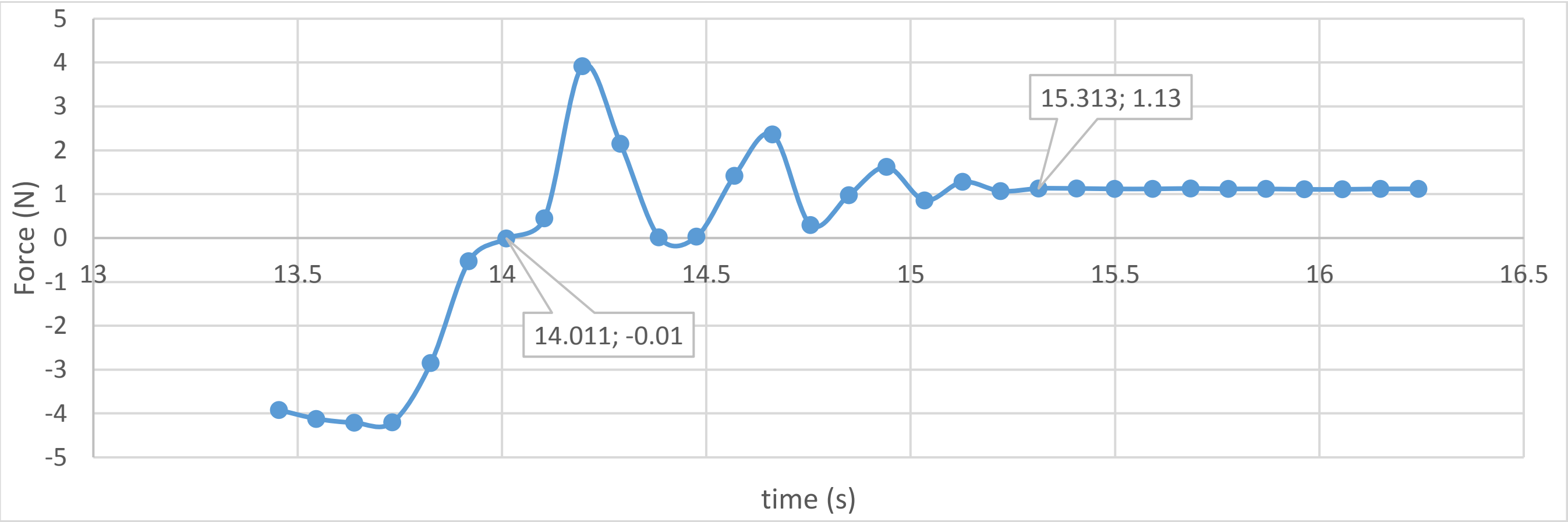

**Fig. 8**. The plot that gives the "force vs. time" data during the motion of the sphere attached to a rope, read from the force sensor.

The values in the figure show the tension in the rope. The net force experienced by the sphere during its motion to be analyzed is the difference between its weight mg=1.12 N and *T*, the tension in the rope. Obviously the net force acting on the sphere varies depending on the variation in the tension. In order to have extensive analyses the temporal variation in the net force acting on the sphere during the motion is shown in Fig. 9.

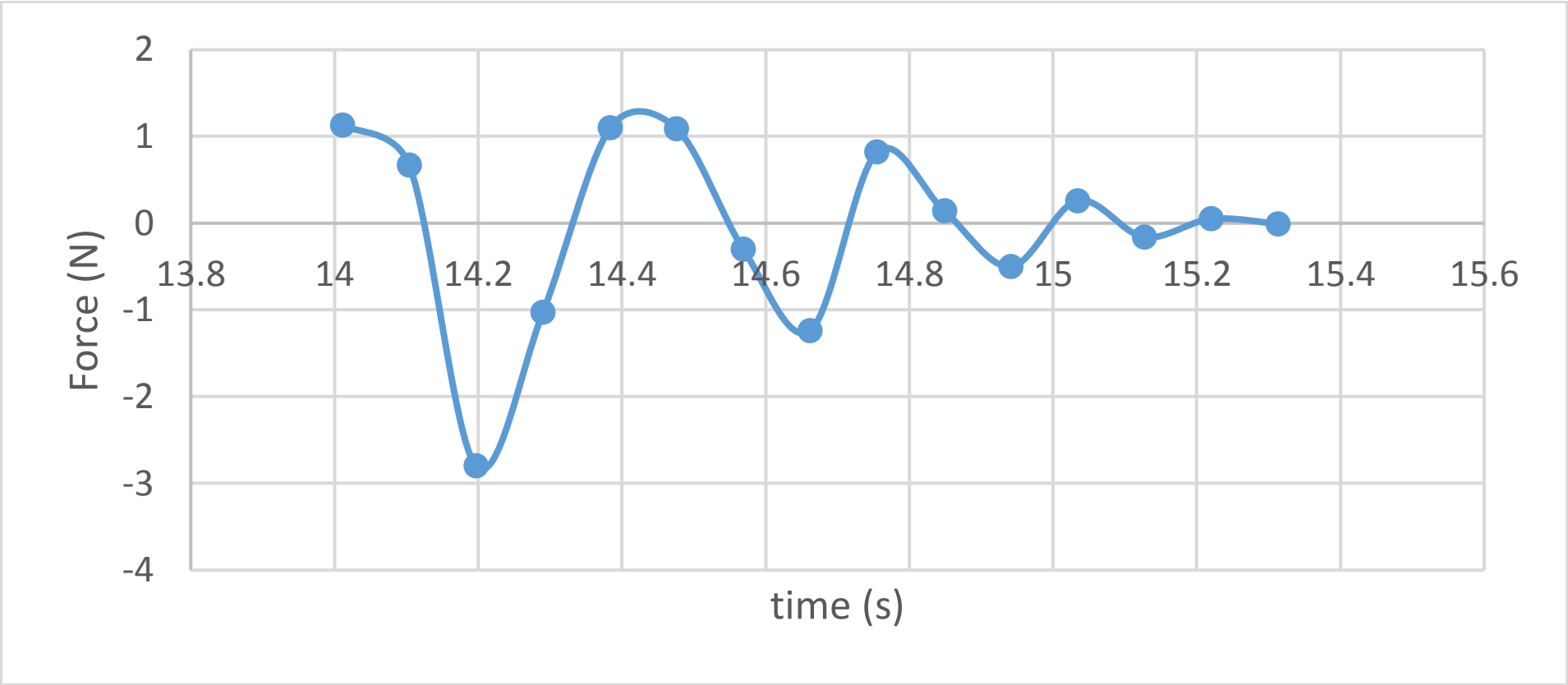

**Fig. 9.** The temporal variation in the net force acting on the sphere during the motion to be analyzed.

It is aimed to calculate the total impulse using the data collected. At this stage of the study the area under the curve of force vs. time graph in Fig. 9 is calculated by Eq. 10 given in the theoretical part. It is found that the total impulse during the motion is Ns. $\sum \vec{I}$=-0.13 Ns.

Since the sphere starts from rest, moves and comes to rest again the net change in its momentum during the whole motion is $\Delta\vec{P}$=0. It is known that theoretically the impulse imparted on the sphere is equal to the change in its momentum. The impulse value obtained as a result of the experimental applications is compatible with theory with a discrepancy of 13%.

## 5. Conclusion

In the study, a single STEM material was developed in accordance with 21st century competencies, such as technological tools and engineering advancements, in order to conduct three different experiments within the framework of a Physics course. At the end of the analysis, it is observed that the acquired results are highly sensitive and consistent with the theoretical values. In light of these characteristics, this application is essential for both inspiring future research and creating instructional resources. Some experiment sets are expensive and cannot be utilized at all educational levels due to funding constraints. However, in this study three different experiments are carried out at a cost of no more than 15 USD and some new experiments can easily be added. For instance the simple pendulum can be used with a nail placed along the vertical line passing through the equilibrium position so that the length of the pendulum changes during the motion. Also by using the first part of the applications of the model "sphere attached to a rope" some free fall analyses (e.g. gravitational acceleration, conservation of energy) can be conducted. In addition to these, unlike the activities to be carried out by some ready-to-use sets students can reach some achievements related with all areas of STEM. Therefore, it seems quite exciting to bring such studies to the classes. The accuracy of the measurements carried out by an Arduino is also noteworthy. Unlike the more conventional measuring devices one can get very sensitive data by the electronic sensors in the Arduino in very short time intervals. For instance the theoretical and directly measured values in the simple pendulum are almost coincident. It is also important in laboratory applications to carry out a precise "measurement / calculation" process when there is a gap of 0.03 N between the maximum and minimum values of the force as far as the measurement precision is concerned and it is easy by the results obtained to see how effective the Arduino is. The third experiment analyzed a process that took a total of 1.302 seconds, and the observed error was found to be quite close to the predicted value. Taking such exact measurements, made possible by the ability to print data at 0.093 second intervals, could usher in a new era of physics experiments being conducted in secondary schools and universities.

## 6.References

## Appendix

*Code used for determine the force chancing by time*

```
#include "HX711.h"
const int DTpin = 3;
const int SCKpin = 2;
HX711 force;
void setup() {
  Serial.begin(9600);
    force.begin(DTpin, SCKpin);}
void loop() {
float F=force.read();
float Ft=((((F-158219.00)/19324.00)*50)/1000.00)*9.81;
{Serial.print(millis()/1000.00,3);
Serial.print("*");
Serial.println(Ft);}}
```